\documentclass[aps,twocolumn,showpacs]{revtex4}
\usepackage{amsmath}
\usepackage{epsfig}
\usepackage{psfrag}

\renewcommand{\!}{\negthickspace}

\begin{document}

\title{Effects of short-range interactions on transport
  through quantum point contacts:\\A numerical approach}

\author{Andreas Lassl}
\author{Peter Schlagheck}
\author{Klaus Richter}
\affiliation{Institut f\"ur Theoretische Physik, Universit\"at
Regensburg, 93040 Regensburg, Germany}

\date{\today}

\begin{abstract}
   We study electronic transport through a quantum point contact, where
   the interaction between the electrons is approximated by a contact
   potential. Our numerical approach is based on the non-equilibrium Green
   function technique which is evaluated at Hartree-Fock level. We show that
   this approach allows us to reproduce relevant features of the so-called ``0.7
   anomaly" observed in the conductance at low temperatures, including the
   characteristic features in recent shot noise measurements. This is consistent
   with a spin-splitting interpretation of the process, and indicates that
   the ``0.7 anomaly" should also be observable in transport experiments with
   ultracold fermionic atoms. 
\end{abstract}


\maketitle


\section{Introduction}

One of the most prominent quantum phenomena in mesoscopic physics is the effect of
conductance quantization. The conductance of a quantum point contact measured as
a function of an applied gate voltage exhibits plateaus at integer multiples of
the conductance quantum, $G_0 = 2e^2/h$, where $-e$ is the electron charge
and $h$ is Planck's constant \cite{Wee88, Wha88, Khu88}. These steps are well
understood in terms of non-interacting electrons \cite{Bee91}. But experimental
conductance curves frequently show an additional plateau-like feature below the first 
conductance step at a value around $0.7\times G_0$. This so-called {\em ``0.7
anomaly"} was first investigated experimentally by Thomas {\em et
al.}~\cite{Tho96} who particularly looked at the magnetic field- and 
temperature dependence of the additional plateau. They found that the
0.7-feature develops smoothly into the Zeeman spin-split plateau at $0.5\times
G_0$ by applying a parallel in-plane magnetic field. That is why those authors
related this anomaly to the spin degree of freedom of the electrons. They
conjectured the presence of spin polarization in quasi one-dimensional
junctions. In addition, Thomas and coworkers revealed that the 0.7 plateau
becomes more pronounced if the temperature is increased. 

Since those first measurements there has been much experimental 
\cite{Tho98, Kri00, Nut00, Rei01, Cro02, Rei02, Roc04, Rok06, Dic06} 
and theoretical 
\cite{Wan96, Bru01, Ber02, Mei02, Hir03, See03, Sta03, Ber05, Rei05, Rei06,
Gol06, Rej06, Ber06} 
effort to explain the origin of this effect. However, a complete
understanding is still missing. Experiments show a zero-bias peak in the
differential conductance typical for the Kondo effect \cite{Cro02}. Furthermore,
the temperature dependence can be characterized by a single parameter which was
interpreted as the Kondo temperature. In a recent experiment \cite{Rok06} a
static spin polarization was measured, which contradicts the Kondo
interpretation. Shot noise measurements \cite{Roc04, Dic06} could show that
two differently transmitting channels contribute to transport.

Theoretical studies of this phenomenon are, on the one hand, based on
calculations using density functional theory (DFT), \cite{Ber06, Wan96, Ber02,
Sta03, Ber05}. In an early publication Wang and Berggren showed how
Coulomb interaction can split the energy levels of up- and down electrons in a
quasi one-dimensional system \cite{Wan96}. They used DFT calculations with 
Hartree- and exchange potentials in local density approximation. 
Their findings were confirmed by more sophisticated calculations which include
exchange-correlation potentials and take into account realistic gate potentials
\cite{Ber02, Sta03, Ber05}. The observed difference of the up-
and down energy levels gives rise to spin dependent transmissions which
manifest themselves in a 0.7 feature in the total conductance. However, to our
knowledge there are no DFT results showing the correct temperature dependence.

Besides DFT calculations, there are various theoretical models describing
different aspects of the 0.7 anomaly. Some models are based on the presence of
spin-splitting \cite{Bru01, Rei05, Rei06}, assuming a density dependent separation
of the up- and down energy levels from the beginning. These models
can qualitatively reproduce the correct magnetic field and temperature
dependence of the 0.7 structure and are also suitable to describe shot noise
\cite{Dic06}. In a complementary approach the 0.7 anomaly is related to the
Kondo effect \cite{Mei02, Hir03} by treating the quantum point contact as
an interacting two-level system for the different spins. Qualitatively, this
approach also leads to the observed temperature and magnetic field behavior of 
the 0.7 feature \cite{Cro02}. Very recently, also shot noise was calculated
within this model showing agreement with experimental data \cite{Gol06}.
Furthermore, interaction with phonons is used to explain the unusual temperature
dependence \cite{See03}. 
 
In this work we present a comparatively simplified approach to the problem,
which is based on the non-equilibrium Green function technique where the 
interaction is incorporated at the Hartree-Fock level. 
We shall, furthermore, approximate the screened Coulomb interaction between
the electrons by a repulsive contact potential.
The fact that we can, within this approach, reproduce all relevant features of
the 0.7 anomaly at temperatures close to zero, including the recently observed
modification of the shot noise factor \cite{Dic06}, supports
arguments in favor of the spin-splitting mechanism \cite{Wan96, Ber02, Sta03,
Ber05, Bru01, Rei05, Rei06}
(in line with experimental evidence provided in Ref.~\cite{Rok06}) and
indicates that the effect is rather robust with respect to the precise
theoretical description of the process.
The short-range potential is furthermore chosen with regard to possible future
transport experiments of ultracold \emph{fermionic atoms} which precisely 
interact via the contact potential that we are using.

This paper is organized as follows: in Section \ref{sec:model} we introduce our
model and present the relevant expressions that are used to calculate the
transport properties. We show our numerical results in Section \ref{sec:results} where we
concentrate on the influence of the coupling constant and the magnetic field on
the conductance. We discuss the zero field case and show results
concerning shot noise and finite temperatures. In Section \ref{sec:disc} we
summarize our results and discuss transport of fermionic atoms through a
constriction. The Appendix consists of a part about determining the strength of
the interaction constant and other model parameters in an ideal two-dimensional
electron gas. Another part contains a detailed description of how to extend the
recursive Green function algorithm to non-equilibrium processes.


\section{The Model}
\label{sec:model}

We describe a two-dimensional electron system with an additional 
in-plane confinement potential $V_{\rm conf}(x, y)$ that defines the geometry of
the quantum point contact. The in-plane magnetic field $\vec{B}=(B,0,0)$
oriented towards the transport direction gives rise to a Zeeman term only. For
moderate magnetic fields the orbital contribution vanishes with the 
choice $\vec{A} = (0, -Bz, 0)$ for the vector potential and $z=0$ for the
location of the two-dimensional electron gas (2DEG). Therefore, the
non-interacting part of the Hamiltonian of the system can be written as  
\begin{equation}
  \label{Hamiltonian}
  H_0^{\sigma} = \frac{p_x^2+p_y^2}{2m} + V_{\rm conf}(x,y) +  g\mu_BB\sigma,
\end{equation}
where $m$ is the effective mass and $\sigma=\pm 1/2$ is the spin quantum number.
The spin-up and spin-down energy levels are separated by the Zeeman energy $E_Z
= g\mu_BB$, where $g$ is the effective gyro-magnetic ratio and $\mu_B$ the
Bohr magneton. Within our model the interaction of two particles located at
$\vec{r}$ and $\vec{r}\,'$ is described by
\begin{equation}
  \label{H_int}
  V_{\rm int}(\vec{r},\vec{r}\,') = \gamma\,\delta(\vec{r}-\vec{r}\,'),
\end{equation}
with interaction strength $\gamma$. This choice of the interaction can
be interpreted as a simple model for an efficiently screened Coulomb
potential. For a homogeneous 2DEG the Thomas-Fermi screening length is of the
order $\lambda_s\approx 5\;\rm nm$ and the width of a typical constriction is
roughly $W\approx 20\;\rm nm$, see Appendix \ref{app:screening}. Therefore, we do
not expect that our model gives an accurate description of the interaction. 
But it provides a transparent physical picture of the mechanism causing spin
splitting. 
Moreover, our Hamiltonian is particularly devised to predict transport features
of neutral fermionic 
atoms, as discussed in section \ref{sec:disc}. In that case it is a very good
approximation to use delta-like interactions.

The coupling constant $\gamma$ can be estimated by calculating the total
interaction energy for a screened Coulomb potential in Thomas-Fermi
approximation, as done in Appendix \ref{app:screening}. We find that
$\gamma\simeq 2\pi\times\hbar^2/(2m)$ gives a realistic order of magnitude for
the interaction strength.

To calculate the transport properties of the system we use the Keldysh Green
function approach \cite{Dat95, Ram86}. This approach is very general as it
allows to treat interactions and to include finite temperatures and bias
voltages. The physical properties are obtained from the retarded and lesser
Green function, $G^r$ and $G^<$. The former can be used to calculate properties
such as the conductance, see Eq.~(\ref{conductance}); from the latter we get the
particle density, see Eq.~(\ref{density}).
Within the Green function approach the interaction is treated in a
self-consistent way and can be included via a proper self-energy. For our
calculations we take into account the first order of the perturbation expansion.
The corresponding retarded self-energy is usually written as a sum of the
Hartree and Fock self-energies, $\Sigma^{\sigma}_H$ and $\Sigma^{\sigma}_F$, and
has the form 
\cite{Dat95} 
\begin{eqnarray}
  \label{selfenergies}
  \Sigma^{\sigma}_H(\vec{r},\vec{r}\,') & = & \frac{-\rm i}{2\pi}\;
  	\delta(\vec{r}- \vec{r}\,')
  	\sum_{\tilde{\sigma}}\int\!{\rm d}\vec{r}\,''\!\int\!{\rm d}E\; 
	V_{\rm int}(\vec{r},  \vec{r}\,'')\times \nonumber\\
  	& & \times\,G^<_{\tilde{\sigma}}(\vec{r}\,''\!,  \vec{r}\,''\!, E),\\
  \Sigma^{\sigma}_F(\vec{r},\vec{r}\,') & = & \frac{\rm i}{2\pi}\int{\rm d}E\; 
  	V_{\rm int}(\vec{r},  \vec{r}\,')\;
	G^<_{\sigma}(\vec{r},  \vec{r}\,'\!, E).\nonumber
\end{eqnarray}
In general, the Hartree self-energy is local, and it involves a sum
over all spin directions, whereas the Fock self-energy is non-local and depends
only on the lesser Green function of the same spin orientation. However, in our
case of delta-interactions, Eq.~(\ref{H_int}), both the Hartree and the Fock
contribution are local. $\Sigma^{\sigma}_F$ exactly
compensates the $(\sigma=\tilde{\sigma})$-term of the spin sum in
$\Sigma^{\sigma}_H$, and we easily obtain for the total interaction self-energy
\begin{equation}
  \label{int_self_energy}
  \Sigma^{\sigma}_{\rm int}(\vec{r}) = \Sigma^{\sigma}_H + \Sigma^{\sigma}_F 
  = \gamma\; n_{-\sigma}(\vec{r}).
\end{equation}
Here, 
\begin{equation}
  \label{density}
  n_{\sigma}(\vec{r}) = -\frac{\rm i}{2\pi}\int{\rm d}E\;
  G^<_{\sigma}(\vec{r}, \vec{r}, E)
\end{equation}
is the density of electrons with spin $\sigma$. 

The Hamilton operator of the interacting system, $\mathcal{H}^{\sigma} =
H^{\sigma}_0 + \Sigma^{\sigma}_{\rm int}(\vec{r})$, is a sum of the
non-interacting Hamiltonian (\ref{Hamiltonian}) and the interaction self-energy
(\ref{int_self_energy}), which acts like an additional local potential. This
potential is different for the different spin directions: a spin-up electron
encounters a potential which is proportional to the density of spin-down
electrons, and vice versa. Hence, there is a repulsive interaction only between
particles with opposite spin directions. Therefore, any small imbalance between
the density of up- and down electrons is increased by this kind of interaction. 

\begin{figure}
  \epsfig{file=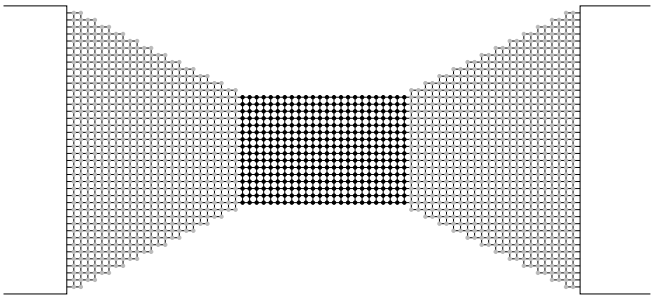, width=\linewidth}
  \caption{The geometry of the system. Each point corresponds to one site of
  the numerical grid; the grey points indicate the region where the interaction
  is gradually switched on and off. The leads are coupled to the left and to the
  right as indicated by the white bars.}
  \label{fig:scatterer}
\end{figure}

To solve the transport problem we discretize the spatial coordinates. The
derivatives in the Hamiltonian are then written as finite differences, and the
Hamilton operator is represented by a block-diagonal matrix \cite{Fru04}. The
diagonal matrix elements contain an on-site energy and all local potentials
$\mathcal{H}^{\sigma}_{ii} = 4\hbar^2/(2ma^2) + V_{\rm conf}(\vec{r}_i) +
\Sigma_{\rm int}(\vec{r}_i) + g\mu_BB\sigma$ (with $a$ the lattice constant).
The off-diagonal matrix elements for neighboring sites $i$ and $j$ are
$\mathcal{H}_{ij} = -\hbar^2/(2ma^2)$ and zero otherwise. As shown in
Fig.~\ref{fig:scatterer}, the geometry of our system has the shape of a linear
constriction with a hard-wall confinement potential $V_{\rm conf}(x,y)$ which 
is zero inside the scattering region, and infinite outside. 

Moreover, the system is coupled to semi-infinite leads that have the same width
as the outer slices of the constriction. The leads are in thermal equilibrium
characterized by the chemical potential $\mu$, and there is no effective
electron-electron interaction in the leads. The interaction potential is
gradually switched on/off in the narrowing region indicated by the grey points
in Fig.~\ref{fig:scatterer}. The coupling to the leads can be exactly taken into
account by self-energies $\Sigma_L$ and $\Sigma_R$ for the left and right lead,
respectively \cite{Dat95, Fer97}. With these ingredients it is possible to
calculate the full retarded Green function by matrix inversion
\begin{equation}
  \label{greenfunction}
  G_{\sigma}^r(E) = \left[E - H^{\sigma}_0 - \Sigma^{\sigma}_{\rm int} -
  \Sigma_L  - \Sigma_R \right]^{-1},
\end{equation}
where $H^{\sigma}_0 $ is given in Eq.~(\ref{Hamiltonian}). The Green function is
a matrix of dimension $2N\times 2N$, where $N$ is the number of lattice sites.
Hence it would be very time consuming to invert the complete matrix in one step,
as the computing time scales like $N^3$. However, it is possible to implement a
recursive algorithm that calculates the Green function of single slices of the
scattering region and couples the slices via a Dyson equation. The details of
this algorithm are explained in Appendix \ref{app:rec_alg}. The recursive scheme
scales with the third power of the width and only linearly with the length of
the system. Thereby it is much more efficient than a direct matrix inversion.

From the retarded Green function we get the lesser function using the kinetic
equation
\begin{equation}
  \label{Glesser}
  G^<_{\sigma}(E) = G_{\sigma}^r(E)\;\Sigma^<(E)\;G_{\sigma}^a(E),
\end{equation}
where the advanced Green function is obtained by hermitian conjugation, 
$G_{\sigma}^a=[G_{\sigma}^r]^{\dagger}$. The lesser self-energy is $\Sigma^<(E)
= -2{\rm i}\, f(E,\mu)\, {\rm Im}(\Sigma_L+\Sigma_R)$, where $f(E,\mu)$ is the
Fermi-Dirac function. This relation holds as the leads are assumed to be in
thermal equilibrium. So the lesser self-energy can be interpreted as the
in-scattering rate for particles with energy $E$ at a chemical potential $\mu$.
The lesser Green function $G^<_{\sigma}(E)$ determines the particle density and
hence the interaction self-energy, according to Eqs.~(\ref{int_self_energy}) and
(\ref{density}). Thus, the interaction self-energy can be calculated from the
retarded Green function, but in turn the retarded Green function depends on the
interaction self-energy. Hence, Eqs.~(\ref{int_self_energy}) and
(\ref{greenfunction}) have to be solved simultaneously. 

The solution is carried out in an iterative way: we start with an initial guess
for the interaction self-energy to calculate the retarded Green function with
Eq.~(\ref{greenfunction}). From this we get the lesser Green function,
Eq.~(\ref{Glesser}), and combining Eqs.~(\ref{int_self_energy}) and
(\ref{density}) we obtain a new value for the interaction self-energy. We
continue with this scheme until we have reached convergence. As soon as we have
found a self-consistent solution we can calculate the conductance of the system
using the Landauer formula
\begin{equation}
  \label{conductance}
  G = \frac{2e^2}{h}{\rm Tr}\Bigl\{\Gamma_L G^r\; \Gamma_R G^a\Bigr\},
\end{equation}
with $\Gamma_{L/R} = {\rm i}\left(\Sigma_{L/R}-\Sigma_{L/R}^{\dagger}\right)$.

\section{Numerical Results}
\label{sec:results}

\subsection{Dependence on the coupling strength}

We first calculate the conductance of the previously described model for zero
temperature. It is convenient to use as the energy unit
$E_1=\hbar^2\pi^2/(2mW^2)$, the energy of the first transverse mode in the
narrow region of the scatterer of width $W$. To break the symmetry between
electrons with different spins we apply a small magnetic field so that the
Zeeman energy has a value $E_Z = g\mu_BB=0.0015\; E_1$. The case of zero
magnetic field is discussed separately in section \ref{sec:zerofield}.

\begin{figure}
  \epsfig{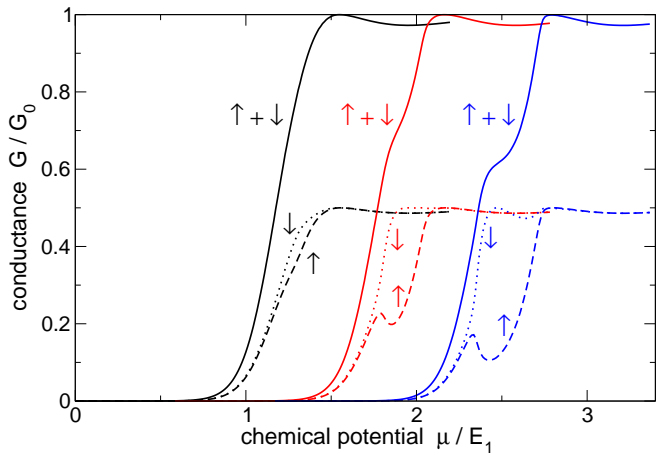}
  \caption{(Color online) Total conductance (solid line) and the up- (dashed)
  and down (dotted) contributions for different interaction constants $\gamma$.
  The coupling constant takes values $\gamma=3.7$ (black), $\gamma=4.1$ (red), 
  and $\gamma=4.5$ (blue) from left to right in units of $\hbar^2/(2m)$. The
  curves for $\gamma\ge 4.1$ have been horizontally offset for clarity.}
  \label{fig:g_dep}
\end{figure}

The conductance for different interaction strengths $\gamma$ is shown in
Fig.~\ref{fig:g_dep}. We find that for a small coupling constant $\gamma=3.7
\times \hbar^2/(2m)$ the up- and down contribution $G_\uparrow$ and
$G_\downarrow$ differ from each other.
This difference is not due to the Zeeman shift, as the Zeeman energy is
approximately two orders of magnitude lower. It is caused by the effective
repulsive interaction between electrons with different spin orientations. If the
interaction strength is increased the up- and down contributions split more and
more. Additionally, a small shoulder develops in the curve for the total
conductance at values between $0.6$ and $0.7$ of the conductance quantum.
This is in agreement with experimental results for the 0.7 anomaly in
Ref.~\cite{Tho96}. For sufficiently high interaction constants the contribution 
of one spin component to the conductance drops down while increasing the 
chemical potential. These spin-resolved conductance curves coincide with results
obtained from transmission across a saddle potential in the presence of a
Gaussian spin-splitting \cite{Nut00}, and with corresponding DFT results
\cite{Ber02}. 

\begin{figure}[b!]
  \epsfig{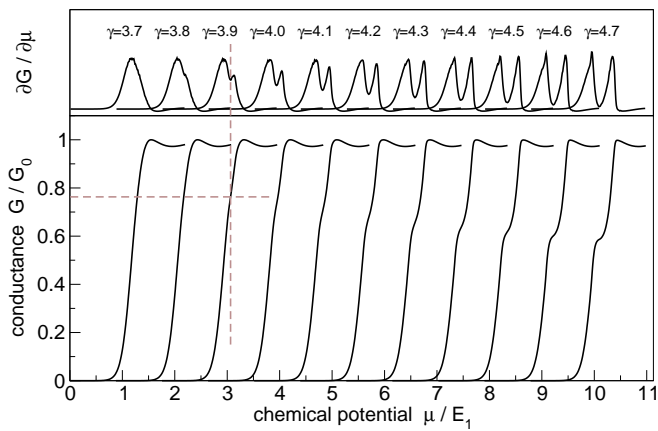}
  \caption{Lower panel: conductance curves for different values of
  the coupling constant ($\gamma$ is given in units of $\hbar^2/(2m)$). The
  derivative (in arbitrary units) of each curve is plotted straight above the
  particular graph. The dashed lines mark the conductance value of the
  ``plateau" for the lowest value of $\gamma$, where a local minimum in
  $\partial G/\partial\mu$ is visible. The different curves
  are offset horizontally.}
  \label{fig:why07}
\end{figure}

The curves in Fig.~\ref{fig:g_dep} already indicate that the spin-splitting has
to be sufficiently strong in order to get a visible effect in the total
conductance. 
This is in accordance with the experimental
results in Ref.~\cite{Rok06} where the authors measure spin resolved
contributions to the total conductance of a point contact. They find that even
in samples that 
are not exhibiting a 0.7 feature the spin-up and -down electrons contribute
differently to the total conductance. Fig.~\ref{fig:why07} shows how the 0.7
plateau develops upon increasing the coupling strength $\gamma$. The lower
panel shows the conductance curves and the upper panel the corresponding
derivatives $\partial G/\partial\mu$. The derivatives change from a 
single peak to a double peak shape as the 0.7 plateau develops. The second peak
in the derivative appears at a coupling constant $\gamma\approx
3.9\times\hbar^2/(2m)$; the corresponding value for the plateau is
$G\approx 0.76\;G_0$. Increasing the interaction constant, the plateau gets more
and more pronounced and eventually converges towards $G_0/2$.
Hence, in our model the interaction parameter $\gamma$
governs the position and the width of the 0.7 plateau. The higher the
conductance value at the plateau, the smaller is its width. There is no plateau
above $0.76\;G_0$ in our model. 

\begin{figure}[b]
  \epsfig{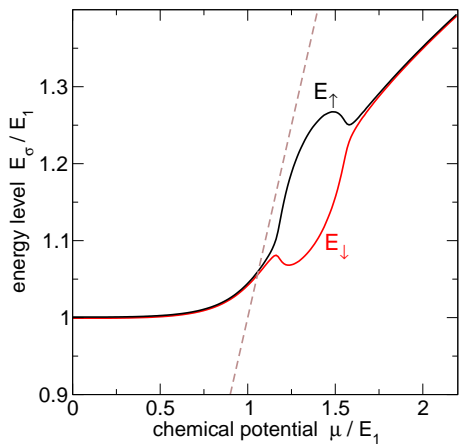}
  \caption{(Color online) Energy levels $E_{\uparrow}$ (black) and
  $E_{\downarrow}$ (red) for $\gamma=4.5\times\hbar^2/(2m)$ and
  $E_Z=0.0015\;E_1$. The dashed curve shows the chemical potential $\mu$.}
  \label{fig:splitting}
\end{figure}

To get information about the energy levels of the different spin orientations we
calculate $E_{\sigma} = E_1^{\sigma} + \langle \Sigma^{\sigma}_{\rm int}
\rangle$. The energy levels of non-interacting electrons $E_1^{\sigma} = E_1 +
g\mu_BB\sigma$ are shifted by the average interaction potential $\langle
\Sigma^{\sigma}_{\rm int} \rangle$ felt by a particle with spin $\sigma$. The
brackets $\langle\ldots\rangle$ denote a spatial average over the region with
full electron-electron interaction (black points in Fig.~\ref{fig:scatterer}).
The resulting curves in Fig.~\ref{fig:splitting} show that the energy levels
$E_{\sigma}$ are located around $E_1$ for $\mu=0$. They are very weakly split by
the Zeeman energy. With increasing $\mu$ the levels rise in energy as the
constriction is populated with electrons and then start to split distinctly, as
soon as the chemical potential is comparable with the energy levels $E_\sigma$.
The reason is that due to the Zeeman splitting the down-level is populated
already at a lower chemical potential causing an imbalance between the density
of spin-up and spin-down carriers in the constriction. The repulsive
interaction between opposite spins tends to increase any imbalance. A small
excess of down electrons repels up electrons from the constriction, which
results in a larger excess of down particles.

The spin-splitting vanishes when the chemical potential is well above both energy
levels. In the range of the chemical potential where the up- and down energy
levels are split, also the up- and down contribution to the conductance differs,
as shown in Fig.~\ref{fig:g_dep}. The obtained energy levels shown in
Fig.~\ref{fig:splitting} are in line with DFT results \cite{Ber02}.

Before comparing with the spin-splitting models \cite{Bru01, Rei05} we shall
note that the quantities plotted in Fig.~\ref{fig:splitting} are only
estimations for the energy levels. Due to the geometry of our system the
transverse modes are broadened with a width of the order of $E_1/2$, as can be
seen in the conductance curves of Fig.~\ref{fig:g_dep}. Therefore, our results
seem to confirm the assumption of the spin-splitting models, that the energy
levels start to split as soon as the chemical potential crosses the up- and down
energy levels. Additionally, we observe a ``pinning" of the upper energy level
to the chemical potential within a substantial range of $\mu$, that means
$E_\uparrow$ evolves parallel to the chemical potential $\mu$ right after the
splitting. The presence of this level pinning is essential in the spin-splitting
models in order to get a 0.7-plateau. In our calculations the plateau also
appears in the range where level pinning is present.

\subsection{Magnetic field dependence}

\begin{figure}
  \epsfig{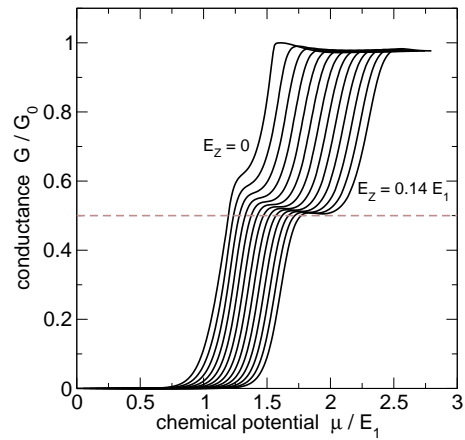}
  \caption{Conductance curves for $\gamma=4.5\times\hbar^2/(2m)$ and different
  magnetic fields. The corresponding Zeeman energies vary from $0$ to $0.14\;E_1$
  in steps of $0.012\;E_1$ from left to right. The curves are horizontally
  offset.}
  \label{fig:Bdep}
\end{figure}

The shape of the conductance curves is influenced by the magnetic field.
Fig.~\ref{fig:Bdep} shows that the 0.7 plateau evolves from a small shoulder at
$G\approx 0.65\;G_0$ to a wide plateau at $G=0.5\;G_0$ as the magnetic field is
increased. This is in agreement with experiments \cite{Tho96, Rok06}. In the
high-field limit Zeeman splitting is the dominant effect. The energy levels of
the different spins are separated by the Zeeman energy which causes a plateau at
$0.5\;G_0$ even in the case of non-interacting electrons. The reason is that
spin-down electrons contribute to transport at chemical potentials $\mu\gtrsim
E_1-g\mu_BB$, whereas for spin-up electrons $\mu\gtrsim E_1+g\mu_BB$ has to be
fulfilled. For strong magnetic fields the effect of electron-electron
interaction is only to broaden the Zeeman spin-split plateau at one half of the
conductance quantum. 

For a more quantitative comparison between our results and experimental data it
is useful to re-scale our quantities and give the magnetic fields in units of
Tesla. 
Therefore, we have to associate an energy value with $E_1=\hbar^2\pi^2/(2mW^2)$.
If we insert the approximate width $W=\sqrt{2/(\pi n)}$ of a quantum point
contact from Eq.~(\ref{width}) we get
\begin{equation}
  \label{B_Tesla}
  B = \frac{1}{g\mu_B}E_Z \approx \frac{\hbar\pi^3 n}{2ge}\frac{E_Z}{E_1}.
\end{equation}
The maximum field applied in Fig.~\ref{fig:Bdep} then corresponds to $B\approx
5.8\,\rm T$ where we used $g=0.44$ for bulk GaAs and a density of $n=1.8\times
10^{11}\,\rm cm^{-2}$ \cite{Tho96}. This magnetic field value is lower than in
experiments where fields of about $10\,\rm T$ are necessary to get a plateau at
$0.5\;G_0$ \cite{Tho96}.

\subsection{The zero-field case}
\label{sec:zerofield}

For the previous calculations we always applied a finite magnetic field. Due to
this field the energy levels of the electrons with different spins were
separated so that the down state can be populated at smaller chemical potentials
than the up state. That is the reason why up electrons are repelled from
entering the constriction, as down electrons are already present at a lower
chemical potential. So the repulsive interaction between particles with opposite
spins leads to an enhancement of an initially small asymmetry between the
density of up- and down electrons in the constriction.

\begin{figure}
  \epsfig{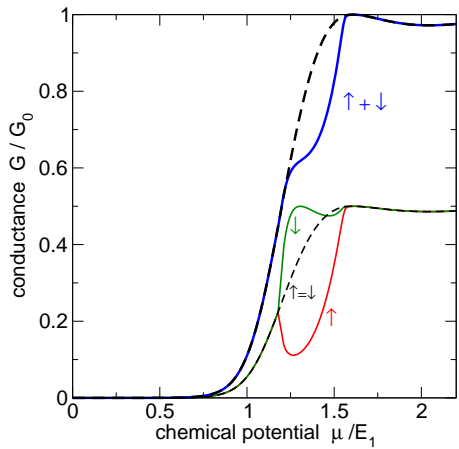}
  \caption{(Color online) Conductance in the zero-field case. The dashed lines
  are for $B=0$; there is no difference between the $\uparrow$ and
  $\downarrow$-contribution and the total conductance does not exhibit a 0.7
  feature. The solid lines show the results for $B\to 0$ where the $\uparrow$
  (red) and $\downarrow$ (green) contributions differ and the total conductance
  (bold blue line) has a shoulder at $G\approx 0.65\;G_0$.} 
  \label{fig:zerofield}
\end{figure}

However, in the case of zero magnetic field the Hamiltonian (\ref{Hamiltonian})
together with the interaction self-energy (\ref{int_self_energy}) is strictly
symmetric with respect to spin-up and spin-down. Therefore, the resulting
conductance curves also have to show the same symmetry. The result for $B=0$ and
$\gamma=4.5\times\hbar^2/(2m)$ is displayed as the dashed curves in
Fig.~\ref{fig:zerofield}. As expected the contributions of the up- and down
electrons exactly coincide and the total conductance has no additional features
below the first step. 

We can investigate the stability of the symmetric solution by slightly
disturbing the symmetry of the system. For each point calculated we start with
a small magnetic field $E_Z^0 = 0.0015\;E_1$. During the first four steps of the
self consistency loop we turn off the magnetic field according to $E_Z^0/s$
where $s$ is the number of the iteration step. After four steps we set the
magnetic field exactly to zero and continue iterating until the results are
converged. In that way we obtain an asymmetric solution for the spin-up and
spin-down contributions. In contrast to the finite field results depicted in
Fig.~\ref{fig:g_dep}, here the splitting sets in abruptly at a chemical
potential $\mu = 1.20\;E_1$. In the range where $G_\uparrow$ and $G_\downarrow$
are different a shoulder appears in the total conductance $G$. Those points
where the spin-splitting is absent coincide with the points for $B=0$. So the
symmetric solution with $G_\uparrow = G_\downarrow$ is unstable and we find a
0.7 anomaly even in the case of zero magnetic field.

In our case the down-contribution to the conductance dominates when we apply a
positive magnetic field. With a negative field the different spin directions
would change their roles. In reality the asymmetry between spin-up and spin-down
may be caused by residual magnetic fields or temporal current fluctuations.
Also magnetic impurities, as well as nuclear spins and dynamic nuclear
polarization might play a role in breaking the up- and down-symmetry. Our
numerical results show that a very weak asymmetry is sufficient to get
spin-splitting. We obtained spin-split results for Zeeman energies down to $E_Z
= 3\times 10^{-7}\;E_1$, corresponding to a magnetic field strength of about
$B\approx 10^{-5}\,\rm T$.

\subsection{Shot noise}

In recent experiments shot noise was measured in quantum point contacts
exhibiting a 0.7 anomaly \cite{Roc04, Dic06}. In the framework of
Landauer-B\"uttiker theory the shot noise power $S$ in a two-terminal device 
is given by \cite{Bla00}
\begin{equation*}
  S = \frac{2e^2}{h}\sum_{n,\sigma}\int{\rm d}E\; 
  T_{n,\sigma}(E)\Big(1-T_{n,\sigma}(E)\Big)\left(f_L-f_R\right)^2.
\end{equation*}
Here, $f_{L/R}$ is the Fermi distribution function of the left/right contact. If
the energy scale on which the transmission functions $T_{n,\sigma}(E)$ vary is
large compared to temperature $k_BT$ and applied source-drain voltage $eV_{\rm
sd}$, the transmissions can be treated as constants. Then the energy integral
over the distribution functions can be performed, yielding
\begin{equation}
  \label{noise}
  S = 2\mathcal{N}\frac{2e^2}{h}\left[eV_{\rm sd}\coth\left( \frac{eV_{\rm
  sd}}{2k_BT} \right) - 2k_BT\right],
\end{equation}
with the noise factor defined as 
\begin{equation}
  \mathcal{N}=\frac{1}{2}\sum_{n,\sigma} T_{n,\sigma}(1-T_{n,\sigma}).
\end{equation}
The noise factor of one single channel vanishes for zero or perfect
transmission, and it is maximal for $T_{n,\sigma}=1/2$.

By simultaneous noise and conductance measurements it is possible to extract
information about spin-resolved transmission coefficients $T_\uparrow$ and
$T_\downarrow$. Whereas the conductance is proportional to the total
transmission, $T_{\rm tot}=T_\uparrow+T_\downarrow$, the noise factor in the
single mode case is $\mathcal{N}=\frac{1}{2}T_{\rm tot}(1-T_{\rm tot}) +
T_\uparrow T_\downarrow$. Only in the case of non-interacting particles where
$T_\uparrow = T_\downarrow = T$ the noise factor reduces to
$\mathcal{N}=T(1-T)$.

\begin{figure}[b]
  \psfrag{noise factor  NF}{\textsf{noise factor} $\mathcal{N}$}
  \epsfig{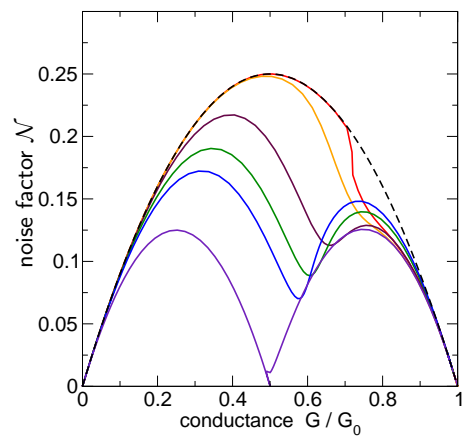}
  \caption{(Color online) Noise factor for $\gamma=4.0\times\hbar^2/(2m)$ and
  for different values of the magnetic field $E_Z/E_1=$ 0.0, 0.0007, 0.006,
  0.012, 0.018, 0.23 from top to bottom. The dashed line indicates the result
  for non-interacting electrons.} 
  \label{fig:noise}
\end{figure}

The authors of Ref.~\cite{Dic06} measured the shot noise power and fitted their
experimental results with Eq.~(\ref{noise}) using $\mathcal{N}$ as fitting
parameter. They find a suppression of noise around the anomalous conductance
plateau. That gives experimental evidence that near the 0.7 feature electrons
are transported by two channels with different transmissions, as also stated in
\cite{Roc04}. This agrees with our results for $G_\uparrow$ and $G_\downarrow$
displayed in Fig.~\ref{fig:g_dep}. 
For conductance values between 0 and 1 the experimentalists find an asymmetric
dome shape for the noise factor evolving into a symmetric double-dome structure
by applying a magnetic field. They are also able to reproduce this behavior
with Reilly's phenomenological model \cite{Rei05}. In a recent publication the
same noise characteristics was obtained using a Kondo model \cite{Gol06}.

The noise factor within our model is depicted in Fig.~\ref{fig:noise} for
$\gamma=4.0\times\hbar^2/(2m)$ and for different magnetic fields. In agreement
with the results in Ref.~\cite{Dic06} we find an asymmetry of the noise factor
with respect to $G=0.5\;G_0$. The shot noise is suppressed at conductance values
around $0.7\;G_0$ which accounts for the differently transmitting channels in
that range. The spin-down channel has almost perfect transparency and hence does
not contribute to the noise factor whereas for $G\sim 0.3\;G_0$ both channels
are equally transmitting and contribute equally to the noise factor. 

By increasing the magnetic field a second maximum appears and the noise factor
evolves towards a symmetric shape. In contrast to the model results in
\cite{Dic06} where the maximum of the right dome is stationary it first rises
slightly in our model and then drops down again for $E_Z/E_1 > 0.04$. 
However, for very strong magnetic fields, $E_Z/E_1 = 0.23$, ($B\approx 9.5\,\rm
T$), the noise factor is symmetric with two maxima at $\mathcal{N}=1/8$. This
accounts for spin resolved transmission of electrons due to the Zeeman
splitting. The discontinuity of the lowest curve at $G\approx G_0/2$ is caused
by the small oscillations of the conductance around $G=G_0$, see
e.g.~Fig.~\ref{fig:g_dep}. In that regime one finds two different noise values
for one conductance value.

\subsection{Temperature dependence}

The 0.7 anomaly is accompanied by a peculiar temperature dependence: within a
certain range the 0.7 plateau gets more pronounced if the temperature is
increased \cite{Tho96, Tho98, Kri00, Cro02}. This behavior can be reproduced by
the spin-splitting models \cite{Bru01, Rei05} as well as by the Kondo model
\cite{Mei02} and by interaction with phonons \cite{See03}. However, to our
knowledge there are no DFT results exhibiting such a temperature behavior
\cite{Ber02}. 

\begin{figure}
  \epsfig{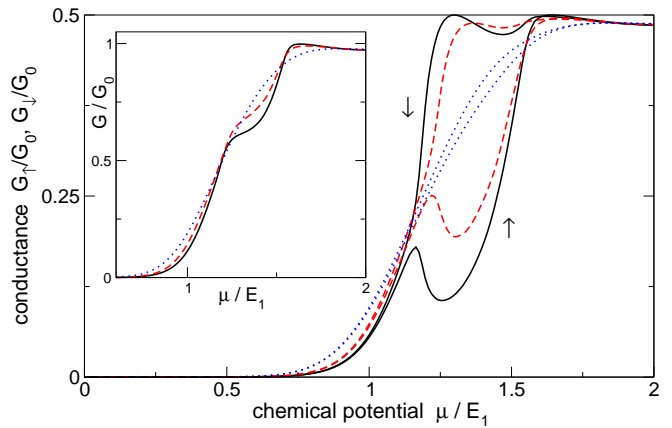}
  \caption{(Color online) The conductance contributions $G_\uparrow$ and
  $G_\downarrow$ for $\gamma=4.5\times\hbar^2/(2m)$ and for different
  temperatures $k_BT=0$ (solid black line), $k_BT=0.029\,E_1$ (dashed red line),
  and $k_BT=0.058\,E_1$ (dotted blue line). The inset shows the corresponding
  total conductance $G=G_\uparrow + G_\downarrow$.} 
  \label{fig:Tdep}
\end{figure}

If we include finite temperatures in our calculations we find a reduction of the
spin-splitting as shown in Fig.~\ref{fig:Tdep}. The difference between the
transmission of up and down electrons is reduced with increasing temperature
which makes the 0.7 plateau less pronounced. This result contradicts the
experimental findings. As DFT calculations are also not able to capture this
phenomenon it is possible that a mean-field description is not sufficient to
explain the temperature dependence. 

In contrast to our approach the spin-splitting models qualitatively yield the
correct temperature dependence \cite{Bru01, Rei05}. In the model approach,
however, the temperature affects only the computation of the conductance whereas
the spin-splitting is assumed to be temperature independent. In our calculation
the temperature also enters in the computation of the density,
Eq.~(\ref{density}), as $G^<(E)$ is truncated around the Fermi level plus
several $k_BT$. Hence, the densities and thus the interaction potentials depend
on the temperature which results in a temperature dependent spin-splitting. The
spin-gap vanishes at temperatures $k_BT\gtrsim 0.08\;E_1$. Second, the
spin-splitting models assume sharp energy levels with step-like transmission
functions $T_{\sigma}(E) = \Theta(E-E_\sigma)$, where $\Theta(x)$ is the
Heaviside step-function. Finite temperatures lead to a smearing of the
conductance and a 0.7 structure is found for temperatures smaller than the
spin-gap, $k_BT<|E_\uparrow-E_\downarrow|$ \cite{Bru01}. In our model the energy
levels exhibit a broadening due to the geometry of the system even at zero
temperature, $k_BT=0$. The broadening is of the order $E_1/2$ (see
Fig.~\ref{fig:g_dep}), larger than the level splitting (see
Fig.~\ref{fig:splitting}). Hence, allowing for finite temperatures the 
broadening is further enhanced which leads to a decrease of the 0.7 plateau.

\section{Discussion and Outlook}
\label{sec:disc}
\subsection{Electron transport}

The presented model describes transport of locally interacting electrons. In
Hartree-Fock approximation only electrons with different spins are interacting
repulsively. This suggests a very intuitive physical picture: if the scattering
region is predominantly occupied by one spin species, electrons with opposite
spin are repelled from the constriction. So this kind of interaction favors an
asymmetric population of the quantum point contact. 
Despite its simplicity the model is adequate to qualitatively explain different
aspects of the 0.7 anomaly. We see how interaction can cause an asymmetry
between the spin-up and spin-down transmission resulting in a shoulder in the
total transmission. The magnetic field dependence of the 0.7 feature is well
reproduced and we find an instability phenomenon in the zero field case leading
to spontaneous spin polarization. Our model also accounts for shot noise
suppression at the 0.7 plateau.

Going beyond Hartree-Fock one expects that spin-splitting is weakened or even
vanishing in the strict one-dimensional case. However, it was shown by exact
methods that Hubbard chain models can have a ferromagnetic ground state if one
is not restricted to exactly one-dimensional systems \cite{Dau98, Kli06}, in
accordance with the Lieb-Mattis theorem \cite{Lie62}. So in our system which is
based on a two-dimensional description a spin polarized ground state is
possible. 

\begin{figure}[b]
  \epsfig{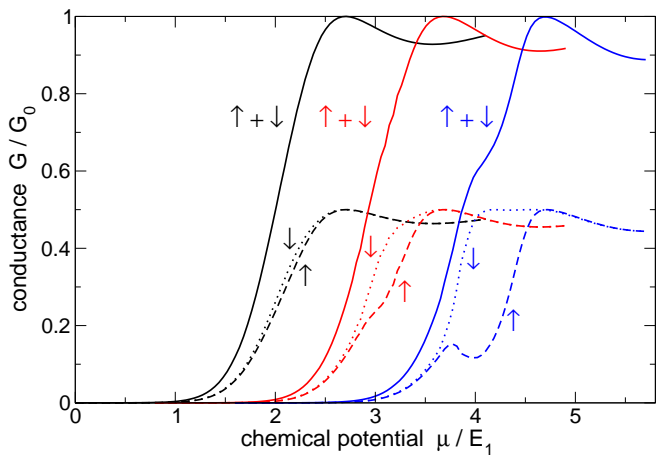}
  \caption{(Color online) Total conductance (solid line) and the up- (dashed) and
  down (dotted) contributions for different interaction constants $\gamma_C$
  calculated with full Coulomb interaction. The dimensionless coupling constant
  takes values $\gamma_C=0.18$ (black), $\gamma_C=0.20$ (red), and
  $\gamma_C=0.22$ (blue) from left to right. The curves for $\gamma_C\ge 0.20$
  have been horizontally offset for clarity.} 
  \label{fig:Coulomb}
\end{figure}

Because the assumption of a delta-like interaction potential might be too crude
for the junction in the 2DEG, we also performed calculations with Coulomb
interaction and full exchange. For the conductance curves we found similar
results as for delta interaction, as shown in Fig.~\ref{fig:Coulomb} for
different coupling strengths. The dimensionless coupling constant can be
estimated as 
\begin{equation}
  \gamma_C = \frac{e^2}{4\pi\varepsilon_0\varepsilon} 
  \frac{2ma^2}{\hbar^2}\frac{1}{a} \approx 0.20,
\end{equation}
where $m=0.07\,m_0$, $\varepsilon=13$ and $a=1\,\rm nm$ was inserted.
Replacing the Coulomb interaction by a Yukawa
potential $V(r)\sim \exp(-r/\lambda)/r$ we find that the results evolve towards
the curves for delta interaction with decreasing screening length $\lambda$. So
the effect of spin-splitting remains robust even in the limit without
screening. In the case of Coulomb interaction the diagonal part of the
Fock self-energy also compensates the short-range contribution to the Hartree
self-energy, Eq.~(\ref{selfenergies}). 
This leads to an effective short-range 
repulsion between different spins, similar to the case of delta-interaction.
We therefore can conclude that the repulsive interaction between electrons with
opposite spin causes spin-splitting even in the case of long-range Coulomb
interaction.

Our approach cannot reproduce the experimentally observed temperature dependence
of the plateau structure \cite{Tho96}, which is also the case for DFT
calculations \cite{Wan96, Ber02, Sta03, Ber05}. This admits two possible
interpretations: On the one hand, Kondo-type correlations could be responsible
for the temperature-induced enhancement of the 0.7 feature.
This mechanism was theoretically suggested in Ref.~\cite{Mei02} and
experimentally supported in Ref.~\cite{Cro02}, but still awaits ultimate
confirmation by \textit{ab initio} calculations that are able to take into
account such correlations and do not involve any tunable parameter.
It was, on the other hand, suggested \cite{See03} that \emph{phonons} could be
at the origin of this effect.

\subsection{Transport of fermionic atoms}

To discriminate between these two complementary interpretations, we propose to
perform transport experiments with \emph{ultracold fermionic atoms}, such as
$^6$Li for instance, which can nowadays be routinely confined within magnetic
or optical trapping potentials and cooled down to temperatures close to the
BCS transition \cite{Zwi05}.
In the context of interaction-induced modifications of the conductance,
\emph{optical} (rather than magnetic) techniques for the confinement of the
atoms would be required in order to trap both spin species of the fermionic
atom.
A quasi two-dimensional configuration could, for instance, be realized by a
rather strong one-dimensional \emph{optical lattice} which creates a sequence
of disk-like confinement geometries for the atoms, and a matter-wave guide
with a constriction could be induced by additional laser beams that are
focused onto the disk within which the atoms are confined.

According to Ref.~\cite{Pet00}, the effective interaction constant $\gamma$ that
characterizes the contact potential (\ref{H_int}) would, in the case of
two-dimensional ultracold fermions, be given by
\begin{equation}
  \gamma \simeq \frac{4 \pi \hbar^2}{m} \frac{1}{\sqrt{2 \pi} \frac{a_\perp}{a_s} + \ln\left(
      \frac{\hbar \omega_\perp}{\pi E} \right)} \, .
\end{equation}
Here, $m$ is the mass of the atom, $\omega_\perp$ denotes the frequency of the harmonic
confinement in the transverse direction (i.e., along the ``third'' dimension),
$a_\perp = \sqrt{\hbar/m\omega_\perp}$ is the corresponding oscillator length, $E$ denotes the
total energy of the collision process between two atoms in the center-of-mass
frame, and $a_s$ represents the $s$-wave scattering length between two atoms
with opposite spin.
Both length scales, $a_s$ and $a_\perp$, can be manipulated, via Feshbach tuning
(see, e.g., Ref.~\cite{Joc03}) as well as through the intensity of
the optical lattice.
It would therefore be possible to realize configurations for which the
effective interaction strength $\gamma$ is of the order of the values that were
discussed in Section \ref{sec:results}.

To measure the atomic 0.7 anomaly, we propose to prepare the fermionic atoms
in a large double-well trap that is optically created within the two-dimensional
confinement geometry, and let them escape from one well to the other through a
small ``bottleneck'' corresponding to the constriction of
Fig.~\ref{fig:scatterer}. Counting the number of atoms that are transported
across the bottleneck within a finite time scale should give rise to a
\emph{current} of atoms close to the Fermi level. This current can be directly
translated into an ``atomic conductance'' in a similar way as in
Ref.~\cite{Thy99}, which would also display a step-like behavior 
when the height of the constriction is lowered by optical techniques.
Magnetic fields can again be used to break the symmetry between spin-up and
spin-down fermions, and the temperature could possibly be controlled by
letting the fermionic cloud interact with a gas or condensate of bosonic atoms
(e.g., by preparing a mixture of $^6$Li and $^7$Li atoms).
As phonons are clearly absent in this setup, any observed feature in the 0.7
anomaly that is not reproducible by mean-field approaches would necessarily be
due to (Kondo-type) correlations.

In short summary, is should be possible to realize transport experiments with
ultracold fermionic atoms where the 0.7 anomaly in the conductance would be
observed.
We expect that such experiments would provide new insight into the central
mechanism that underlies this phenomenon.

\begin{acknowledgments}
We thank Milena Grifoni, Tobias Paul and Michael Wimmer for helpful discussions
and we acknowledge support by the {\em Deutsche Forschungsgemeinschaft}
within the Research Training Group GRK 638.
\end{acknowledgments}


\begin{appendix}
\section{Estimation of the model parameters}
\label{app:screening}

The coupling constant $\gamma$ is the main parameter of our model. It has to be
sufficiently high in order to get an observable effect of the electron-electron
interaction. Here we want to estimate an upper limit of the
interaction strength using an exponentially screened Coulomb potential.

In a homogeneous 2DEG the screening length $\lambda_s$ in Thomas-Fermi
approximation is given by \cite{And82}
\begin{equation}
  \label{scr_length}
  \lambda_s = \frac{2\pi\varepsilon_0\varepsilon\hbar^2}{me^2},
\end{equation}
where $\varepsilon$ is the average dielectric constant of the two materials on
both sides of the 2DEG. For a GaAs/AlGaAs interface we find $\lambda_s\approx
5\,\rm nm$, where $m = 0.07\,m_0$ and $\varepsilon=13$ was used. To compare the
screening length with the width of the constriction, we have to estimate the
typical dimensions of a point contact. The lithographic width is of the order of
several hundred nanometers, but the electrons are confined by the electrostatic
potential due to the gates. The effective width of the constriction is then
controlled by the gate voltage. From experiments we know the typical density of
carriers $n=1.8\times 10^{11}\,\rm cm^{-2}$ which is related to the chemical
potential $\mu = \hbar^2\pi n/m$. When the first channel opens the effective
width can be estimated by equating the chemical potential with the energy of the
first sub-band $\hbar\omega/2$ for a parabolic confinement. The width of the
confinement potential at this energy is $W=2\sqrt{\hbar/(m\omega)}$, which gives
\begin{equation}
  \label{width}
  W=\sqrt{\frac{2}{\pi n}}.
\end{equation}
So we find that the effective width of a quantum point contact is of the order
$W=20\,\rm nm$. Inside the constriction the density is expected to be lower
than in the homogeneous 2DEG, so the effective width will be larger than the
above estimated value.

For $\delta$-interaction the coupling constant is given by the spatial integral
over the interaction Hamiltonian. So we calculate the corresponding quantity for
a screened Coulomb potential
\begin{equation}
  \gamma = \frac{e^2}{4\pi\varepsilon_0\varepsilon}\int{\rm d}^2r'
  \frac{{\rm e}^{-|\vec{r}\,'|/\lambda_s}}{|\vec{r}\,'|}
  = \frac{e^2\lambda_s}{2\varepsilon_0\varepsilon}.
\end{equation}
Inserting the screening length, Eq.~(\ref{scr_length}), we find
\begin{equation}
  \gamma = 2\pi\frac{\hbar^2}{2m}.
\end{equation}
This is just a rough estimation as several aspects are neglected. First, 
in Thomas-Fermi approximation the screening length in two dimensions is
independent of the electron density. But beyond this approximation one finds an
increasing screening length as the charge density goes to zero \cite{And82}.
This reflects that screening is less efficient if the particle density is too
small. 

Second, screening in two dimensions is not as strong as in three-dimensional
systems. The asymptotic behavior of the screened potential is not exponential,
but it follows an $r^{-3}$ law \cite{And82}. However, the resulting coupling
constant does not differ dramatically from the one obtained by exponential
screening. Both facts would give rise to an even higher upper limit of the
coupling strength.

\section{Recursive algorithm for non-equilibrium Green functions}
\label{app:rec_alg}

\begin{figure}
  \epsfig{file=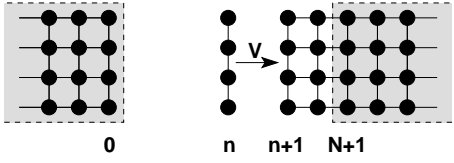, width=6cm}
  \caption{The Green function is constructed by coupling single slices starting
  from one of the leads (grey regions).}
  \label{fig:recursive}
\end{figure}

The recursive Green function algorithm is widely used for calculating electronic
properties of two- and three-dimensional systems. The basic idea is to build up
the full Green function slice by slice instead of evaluating it in one step.
Thus, the dimensions of the matrices that have to be inverted are strongly
reduced. If the Green function $G_0$ of a semi-infinite region and an adjacent
{\em isolated} slice is known, it is possible to calculate the Green function
$G$ of the {\em coupled} system using Dyson's equation
\begin{equation}
  \label{dyson}
  G = G_0 + G_0VG.
\end{equation}
(For this derivation we omit the spin index $\sigma$ and the superscript $r$ for
the retarded functions). 
Here, $V$ denotes the hopping matrix between the two adjacent slices. The Green
function of a semi-infinite lead can be calculated analytically \cite{Fer97}. So
it is  possible to start with an isolated lead and then add slice by slice until
the opposite lead is reached. This is schematically shown in
Fig.~\ref{fig:recursive}. After coupling that lead to the rest of the system one 
has obtained the Green function of the complete system at the surface of one
lead. This Green function contains all information to calculate the current
through the system. The above described procedure is explained for example in 
Refs.~\cite{Fer97, Mac95}. 

As we are interested also in the electronic density which is determined by the
lesser function $G^<$, the above explained algorithm is not sufficient. Here we
present an extension to the usual recursive Green function method which allows
us to calculate the retarded Green function between the two leads as well as the
lesser Green function (see also Ref.~\cite{Lak96}). The condition to apply this
algorithm is that all 
relevant self-energies are diagonal so that the effective Hamiltonian that has
to be inverted, Eq.~(\ref{greenfunction}), keeps its block-diagonal structure.
This condition is fulfilled for the Hartree self-energy and also for the Fock
self-energy in our case of delta interaction. In the general case of full
Coulomb interaction the Fock self-energy is not diagonal, see
Eq.~(\ref{selfenergies}), so the presented method can not be used.

We first show how to add one single slice to a semi-infinite region. In the
following we use the notation $G^{S(n)}$ for the Green function of an isolated
slice $n$, and $G^{R/L(n)}$ for the Green function of the right/left
semi-infinite region starting at slice $n$. The full Green function of the
complete (infinite) system is denoted by $G$ (without superscripts).
In order to couple the Green function $G^{S(n)}$ of the isolated slice $n$ to
the Green function $G^{R(n+1)}$ that covers all lattice sites to the right of
$(n+1)$, we use the Dyson equation (\ref{dyson}),
\begin{eqnarray*}
  \left\langle n \left| G^{R(n)} \right| n \right\rangle & = & 
  \left\langle n \left| \left(G^{S(n)}+G^{R(n+1)}\right)\right| n \right\rangle
  + \\ 
  && \left\langle n \left| \left(G^{S(n)}+G^{R(n+1)}\right)VG^{R(n)}\right| n
  \right\rangle.
\end{eqnarray*}
As $G^{R(n+1)}$ has no matrix elements with slice $n$, the terms $\langle n|
G^{R(n+1)} |n\rangle$ vanish and we get
\begin{eqnarray*}
  \left\langle n \left| G^{R(n)} \right| n \right\rangle & = & 
  \left\langle n \left| G^{S(n)}\right| n \right\rangle  + \\ 
  && \sum_{a,b}\langle n | G^{S(n)}|a\rangle 
  \langle a | V|b\rangle \langle b | G^{R(n)}|n\rangle .
\end{eqnarray*}
Noting that $G^{S(n)}$ has only non-zero matrix elements with slice $n$ we
get the constraint $a=n$. As the coupling matrix $V$ acts only between adjacent
slices and has no overlap with other slices $b$ is restricted to the values
$n\pm 1$. With $\langle n-1|G^{R(n)}|n-1\rangle=0$ we find
\begin{equation}
  \label{Gnn_zw}
  G^{R(n)}_{n,n} = G^{S(n)}_{n,n} + G^{S(n)}_{n,n}V^{}_{n,n+1}G^{R(n)}_{n+1,n},
\end{equation}
where $G_{n,m}=\langle n|G|m\rangle$ is the sub-matrix of $G$ related to the
slices $n$ and $m$. The Green function $G^{R(n)}_{n+1,n}$ appearing in
Eq.~(\ref{Gnn_zw}) can be calculated via the Dyson equation in a similar way,
and we get
\[
  G^{R(n)}_{n+1,n} = G^{R(n+1)}_{n+1,n+1}V^{}_{n+1,n}G^{R(n)}_{n,n}.
\]
Inserting this result into Eq.~(\ref{Gnn_zw}) and solving for $G^{R(n)}_{n,n}$
we obtain
\begin{equation}
  \label{GRnn}
  G^{R(n)}_{n,n} = \left[ (E-H_{n,n}) -
  V^{}_{n,n+1}G^{R(n+1)}_{n+1,n+1}V^{}_{n+1,n} \right]^{-1},
\end{equation}
where we used $G^{S(n)}_{n,n} = (E-H_{n,n})^{-1}$. Therefore,
Eq.~(\ref{GRnn}) allows us to calculate the Green function covering all lattice
sites to the right of slice $n$ from the Green function to the right of $(n+1)$.
In that way we have added one slice. Iterating this scheme we can finally obtain
the Green function $G^{R(1)}_{1,1}$ at the left end of the scattering region.
Then one has to connect the Green functions of the two semi-infinite sections to
get the full Green function $G_{1,1}$ (without superscript) at the left end of
the scatterer. This we obtain by using the Dyson equation
\begin{eqnarray*}
  \left\langle 1 \left| G \right| 1 \right\rangle & = & 
  \left\langle 1 \left| \left(G^{L(0)}+G^{R(1)}\right)\right| 1 \right\rangle
  + \\ 
  && \left\langle 1 \left| \left(G^{L(0)}+G^{R(1)}\right)VG\right| 1
  \right\rangle,
\end{eqnarray*}
and we find
\begin{equation}
  \label{G11}
  G_{1,1} = \left[ 1 - G^{R(1)}_{1,1} V^{}_{1,0} G^{L(0)}_{0,0}
  V^{}_{0,1} \right]^{-1} G^{R(1)}_{1,1}.
\end{equation}
In this equation $G^{L(0)}_{0,0}$ is the surface Green function of the
semi-infinite left lead. The Green function $G_{1,1}$ contains all information
about the reflection coefficients at the left lead.

In an analog way we can start from the left lead and calculate all Green
functions from left to right by
\begin{equation}
  \label{GLnn}
  G^{L(n)}_{n,n} = \left[ (E-H_{n,n}) -
  V^{}_{n,n-1}G^{L(n-1)}_{n-1,n-1}V^{}_{n-1,n} \right]^{-1},
\end{equation}
and finally obtain the full Green function at the right end of the scatterer,
\begin{equation}
  \label{GNN}
  G_{N,N} = \left[ 1 - G^{L(N)}_{N,N} V^{}_{N,N+1} G^{R(N+1)}_{N+1,N+1}
  V^{}_{N+1,N} \right]^{-1} G^{L(N)}_{N,N}.
\end{equation}
Here $G^{R(N+1)}_{N+1,N+1}$ is the surface Green function of the right lead.

Knowing the full Green functions at both ends of the scatterer, $G_{1,1}$ and
$G_{N,N}$, we can now compute the full Green functions between the ends and any
slice $n$ inside the scattering region. We use the Dyson equation
\begin{eqnarray*}
  \left\langle n \left| G \right| 1 \right\rangle & = & 
  \left\langle n \left| \left(G^{L(n-1)}+G^{R(n)}\right)\right| 1 \right\rangle
  + \\ 
  && \left\langle n \left| \left(G^{L(n-1)}+G^{R(n)}\right)VG\right| 1
  \right\rangle
\end{eqnarray*}
to obtain
\begin{equation}
  \label{Gn1}
  G_{n,1} = G^{R(n)}_{n,n} V_{n,n-1}G_{n-1,1},
\end{equation}
where the $G^{R(n)}_{n,n}$ are calculated from Eq.~(\ref{GRnn}). Analogously one
finds 
\begin{equation}
  \label{GnN}
  G_{n,N} = G^{L(n)}_{n,n} V_{n,n+1}G_{n+1,N}
\end{equation}
with the Green functions $G^{L(n)}_{n,n}$ from Eq.~(\ref{GLnn}). The last two
equations allow us to compute the Green functions $G_{n,1}$ and $G_{n,N}$
recursively by starting with the Green functions $G_{1,1}$ and $G_{N,N}$ at the
ends of the scattering region.

Now it is possible to compute the diagonal elements of the lesser Green function
which are needed to calculate the electron density, Eq.~(\ref{density}). A
diagonal matrix element of $G^<$ reads according to Eq.~(\ref{Glesser})
\begin{equation}
  \label{Gless}
  \left[G^<\right]_{xx} = \sum_{i,j} [G]_{xi} \left[\Sigma^<\right]_{ij}
  [G]_{xj}^*
\end{equation}
with $i,j \in \{n=1, n=N\}$.
The self-energy $\Sigma^<$ is only non-zero at the ends of the scatterer where
the lattice sites are coupled to the leads. So the indices $i$ and $j$ are from
the first and last slice of the scattering region. Therefore, the Green
functions calculated from Eqs.~(\ref{Gn1}) and (\ref{GnN}) enter here.

The complete recursive procedure can be summarized in the following steps:
\begin{itemize}
  \item calculate and store all $G^{R(n)}_{n,n}$ from the lead Green function
  $G^{R(N+1)}_{N+1,N+1}$ by means of Eq.~(\ref{GRnn});
  \item calculate and store all $G^{L(n)}_{n,n}$ from the lead Green function
  $G^{L(0)}_{0,0}$ by means of Eq.~(\ref{GLnn});
  \item compute the full Green functions $G_{1,1}$ and $G_{N,N}$ at the left and
  right end of the scatterer using Eqs.~(\ref{G11}) and (\ref{GNN});
  \item use Eqs.~(\ref{Gn1}) and (\ref{GnN}) to calculate and store all
  $G_{n,1}$ and $G_{n,N}$ from $G_{1,1}$ and $G_{N,N}$. One of those Green
  functions is $G_{N,1}$ which contains information about the transport
  properties;
  \item finally, one obtains the diagonal elements of $G^<$ from $G_{n,1}$ and
  $G_{n,N}$ by means of Eq.~(\ref{Gless}).\\
\end{itemize}

In total, one has to run four times through the entire system in order to be
able to calculate $G^<$ as well as parts of $G^r$ which are needed for the
reflection and transmission coefficients. If the calculation of $G^<$  is not
necessary it is enough to pass the system twice to get all reflection and
transmission coefficients. So the scheme reduces to the standard recursive
algorithm \cite{Fer97, Mac95}. If one is only interested in the current, passing
the scatterer once is sufficient. After computing $G_{1,1}$ with Eq.~(\ref{G11})
the total reflection is known. Employing current conservation (unitarity) it is
possible to get the total transmission and hence the current.

\end{appendix}



\begin{thebibliography}{----}
\bibitem{Wee88}
B.~J. van Wees, H. van Houten, C.~W.~J. Beenakker, J.~G. Williamson, L.~P.
Kouwenhoven, D. van der Marel, and C.~T. Foxon, Phys. Rev. Lett. \textbf{60}, 848
(1988)

\bibitem{Wha88}
D.~A. Wharam, T.~J. Thornton, R. Newbury, M. Pepper, H. Ahmed, J.~E.~F. Frost,
D.~G. Hasko, D.~C. Peacock, D.~A. Ritchie, and G.~A.~C. Jones, J. Phys. C
\textbf{21}, L209 (1988)

\bibitem{Khu88}
A. Khurana, Physics Today November 1988, p21 (1988)

\bibitem{Bee91}
for an overview see e.g. C.~W.~J. Beenakker, and H. van Houten, Sol. State Phys.
\textbf{44}, 1 (1991); cond-mat/0412664 

\bibitem{Tho96}
K.~J. Thomas, J.~T. Nicholls, M.~Y. Simmons, M. Pepper, D.~R. Mace, D.~A.
Ritchie, Phys. Rev. Lett. \textbf{77}, 135 (1996)

\bibitem{Tho98}
K.~J. Thomas, J.~T. Nicholls, N.~J. Appleyard, M.~Y. Simmons, M. Pepper, D.~R.
Mace, W.~R. Tribe, and D.~A. Ritchie, Phys. Rev. B \textbf{58}, 4846 (1998)

\bibitem{Kri00}
A. Kristensen, H. Bruus, A.~E. Hansen, J.~B. Jensen, P.~E. Lindelof, C.~J.
Marckmann, J. Nyg\r{a}rd, C.~B. S\o rensen, F. Beuscher, A. Forchel, and M.
Michel, Phys. Rev. B \textbf{62}, 10950 (2000)

\bibitem{Nut00}
S. Nuttinck, K. Hashimoto, S. Miyashita, T. Saku, Y. Yamamoto, and Y. Hirayama,
Jpn. J. Appl. Phys. \textbf{39}, L655 (2000)

\bibitem{Rei01}
D.~J. Reilly, G.~R. Facer, A.~S. Dzurak, B.~E. Kane, R.~G. Clark, P.~J. Stiles,
R.~G. Clark, A.~R. Hamilton, J.~L. O'Brien, N.~E. Lumpkin, L.~N. Pfeiffer, and
K.~W. West, Phys. Rev. B \textbf{63}, 121311(R) (2001)

\bibitem{Cro02}
S.~M. Cronenwett, H.~J. Lynch, D. Goldhaber-Gordon, L.~P. Kouwenhoven, C.~M.
Marcus, K. Hirose, N.~S. Wingreen, and V. Umansky, Phys. Rev. Lett. \textbf{88},
226805 (2002) 

\bibitem{Rei02}
D.~J. Reilly, T.~M. Buehler, J.~L. O'Brien, A.~R. Hamilton, A.~S. Dzurak, R.~G.
Clark, B.~E. Kane, L.~N. Pfeiffer, and K.~W. West, Phys. Rev. Lett. \textbf{89},
246801 (2002) 

\bibitem{Rok06}
L.~P. Rokhinson, L.~N. Pfeiffer, and K.~W. West, Phys. Rev. Lett. \textbf{96},
156602 (2006)

\bibitem{Roc04} 
P. Roche, J. S\'{e}gala, D.~C. Glattli, J.~T. Nicholls, M. Pepper,
A.~C. Graham, K.~J. Thomas, M.~Y. Simmons, and D.~A. Ritchie,  Phys. Rev. Lett.
\textbf{93}, 116602 (2004)

\bibitem{Dic06} 
L. DiCarlo, Y. Zhang, D.~T. McClure, D.~J. Reilly, C.~M. Marcus,
L.~N. Pfeiffer, and K.~W. West, Phys. Rev. Lett. \textbf{97}, 036810 (2006)

\bibitem{Ber06}
for a short recent review see K.-F. Berggren, Turk. J. Phys. \textbf{30}, 197
(2006) 

\bibitem{Wan96}
C.-K. Wang and K.-F. Berggren, Phys. Rev. B \textbf{54}, R14257 (1996)

\bibitem{Ber02}
K.-F. Berggren and I.~I. Yakimenko, Phys. Rev. B \textbf{66}, 085323 (2002)

\bibitem{Sta03}
A.~A. Starikov, I.~I. Yakimenko, and K.-F. Berggren, Phys. Rev. B \textbf{67},
235319 (2003)

\bibitem{Ber05}
K.-F. Berggren, P. Jaksch, and I. Yakimenko, Phys. Rev. B \textbf{71}, 115303 (2005)

\bibitem{Bru01}
H. Bruus, V.~V. Cheianov, and K. Flensberg, Physica E \textbf{10}, 97 (2001)

\bibitem{Rei05}
D.~J. Reilly, Phys. Rev. B \textbf{72}, 033309 (2005)

\bibitem{Rei06}
D.~J. Reilly, Y. Zhang, and L. DiCarlo, Physica E \textbf{34}, 27 (2006)

\bibitem{Mei02}
Y. Meir, K. Hirose, and N.~S. Wingreen, Phys. Rev. Lett. \textbf{89}, 196802 (2002)

\bibitem{Hir03}
K. Hirose, Y. Meir, and N.~S. Wingreen, Phys. Rev. Lett. \textbf{90}, 026804 (2003) 

\bibitem{See03}
G. Seelig and K.~A. Matveev, Phys. Rev. Lett. \textbf{90}, 176804 (2003)

\bibitem{Gol06}
A. Golub, T. Aono, and Y. Meir, cond-mat/0605114 (2006)

\bibitem{Rej06}
T. Rejec and Y. Meir, Nature \textbf{442}, 900 (2006)

\bibitem{Dat95} 
S. Datta: {\em Electronic Transport in Mesoscopic Systems} 
(Cambridge University Press, 1995)

\bibitem{Ram86}
J. Rammer and H. Smith,  Rev. Mod. Phys. \textbf{58}, 323 (1986)

\bibitem{Fru04}
see for example: D. Frustaglia, M. Hentschel, and K. Richter, Phys. Rev. B
\textbf{69}, 155327 (2004) 

\bibitem{Fer97} 
D.~K. Ferry and S.~M. Goodnick: {\em Transport in Nanostructures}
(Cambridge University Press, 1997)

\bibitem{Bla00}
Ya.~M. Blanter and M. B\"uttiker, Phys. Rep. \textbf{336}, 1 (2000)

\bibitem{Dau98}
S. Daul and R.~M. Noack, Phys. Rev. B \textbf{58}, 2635 (1998)

\bibitem{Kli06}
A.~D. Klironomos, J.~S. Meyer, and K.~A. Matveev, Europhys. Lett. \textbf{74}, 679
(2006)

\bibitem{Lie62}
E. Lieb and D. Mattis, Phys. Rev. \textbf{125}, 164 (1962)

\bibitem{Zwi05}
M.~W. Zwierlein, J.~R. Abo-Shaeer, A. Schirotzek, C.~H. Schunck, and W.
Ketterle, Nature \textbf{435}, 1047 (2005).

\bibitem{Pet00}
D.~S. Petrov, M. Holzmann, and G.~V. Shlyapnikov, Phys. Rev. Lett.
\textbf{84}, 2551 (2000).

\bibitem{Joc03}
S. Jochim, M. Bartenstein, A. Altmeyer, G. Hendl, S. Riedl, C. Chin, J. Hecker
Denschlag, and R. Grimm, Science \textbf{302}, 2101 (2003).

\bibitem{Thy99}
J.~H. Thywissen, R.~M. Westervelt, and M. Prentiss, Phys. Rev. Lett.
\textbf{83}, 3762 (1999).

\bibitem{And82}
T. Ando, A.~B. Fowler, and F. Stern, Rev. Mod. Phys. \textbf{54}, 437 (1982);
Note: the authors are using cgs-units.

\bibitem{Mac95}
M. Macucci, A. Galick, and U. Ravaioli, Phys. Rev. B \textbf{52}, 5210 (1995)

\bibitem{Lak96}
R. Lake, G. Klimeck, R.~C. Bowen, and D. Jovanovic, J. Appl. Phys. \textbf{81}, 7845
(1996)

\end{thebibliography}
\end{document}